\documentclass[oldversion]{aa}
\usepackage{graphicx}
\usepackage{ulem}
\usepackage{threeparttable}
      \def\new#1 {{\bf #1 }}
      \def\cut#1 {\sout{#1} }
\begin{document}
\def\ffam {\hbox{$\,.\!\!^{\prime}$}}
\def\ffas {\hbox{$\,.\!\!^{\prime\prime}$}}
\def\ffM {\hbox{$\,.\!\!^{\rm M}$}}
\def\ffm {\hbox{$\,.\!\!^{\rm m}$}}
\def\ffs {\hbox{$\,.\!\!^{\rm s}$}}
\def\kms    {\ifmmode{{\rm \ts km\ts s}^{-1}}\else{\ts km\ts s$^{-1}$}\fi}
\def\msol   {\ifmmode{{\rm M}_{\odot} }\else{M$_{\odot}$}\fi}
\def\lsol   {\ifmmode{L_{\odot}}\else{$L_{\odot}$}\fi}
\def\lfir   {\ifmmode{L_{\rm FIR}}\else{$L_{\rm FIR}$}\fi}
\def\CO32    {$^{12}$CO(3--2) }
\def\13CO32  {$^{13}$CO(3--2) }
\def \Kkmspc{K\,\kms\,pc$^2$}
\def\,{\thinspace}
\def\Msun{M$_\odot$}
\def\msun{M$_\odot$}
\def\Lsun{L$_\odot$}
\def\Lfir{{\hbox {$L_{\rm FIR}$}}}

\title{Weak $^{13}$CO in the Cloverleaf Quasar: evidence for a young,
        early generation starburst}

\titlerunning{$^{13}$CO in the Cloverleaf Quasar}

\author{C. Henkel\inst{1}
        \and
        D. Downes\inst{2}
        \and
        A. Wei{\ss}\inst{1}
        \and
        D. Riechers\inst{3,4,5}
        \and
        F. Walter\inst{4}}

\institute{Max-Planck-Institut f{\"u}r Radioastronomie, Auf dem H\"ugel 69, D-53121 Bonn, Germany
           \and
           Institut de Radio Astronomie Millim{\'e}trique, Domaine
           Universitaire, F-38406 St.-Martin-d'H{\`e}res, France
           \and
           California Institute of Technology, Astronomy Department,
           MC\,249--17, 1200 East California Boulevard, Pasadena, CA
           91125, USA 
           \and
           Max-Planck-Institut f{\"u}r Astronomie, K{\"o}nigstuhl 17,
           D-69117 Heidelberg, Germany
           \and
           Hubble Fellow}

\authorrunning{Henkel et al.}

\offprints{C. Henkel, \email{chenkel@mpifr-bonn.mpg.de}}

\date{Received date / Accepted date}

\abstract{
Observations of $^{12}$CO at high redshift indicate rapid metal enrichment in the nuclear
regions of at least some galaxies in the early universe. However, the enrichment may be 
limited to nuclei that are synthesized by short-lived massive stars, excluding classical 
``secondary'' nuclei like $^{13}$C. Testing this idea, we used the IRAM Interferometer to 
tentatively detect the $^{13}$CO $J$=3$\rightarrow$2 line at a level of 0.3\,Jy\,km\,s$^{-1}$ 
toward the Cloverleaf Quasar at $z$ = 2.5. This is the first observational evidence for 
$^{13}$C at high redshift. The $^{12}$CO/$^{13}$CO $J$=3$\rightarrow$2 luminosity ratio is 
with 40$^{+25}_{-8}$ much higher than ratios observed in molecular clouds of the 
Milky Way and in the ultraluminous galaxy Arp220, but may be similar to that observed toward 
NGC~6240. Large Velocity Gradient models simulating seven $^{12}$CO transitions and the 
$^{13}$CO line yield $^{12}$CO/$^{13}$CO abundance ratios in excess of 100. It is possible 
that the measured ratio is affected by a strong submillimeter radiation field, which reduces 
the contrast between the $^{13}$CO line and the background. It is more likely, however, that 
the ratio is caused by a real deficiency of $^{13}$CO. This is already apparent in local 
ultraluminous galaxies and may be even more severe in the Cloverleaf because of its young age 
($\la$2.5\,Gyr). A potential conflict with optical data, indicating high abundances also 
for secondary nuclei in quasars of high redshift, may be settled if the bulk of the CO emission 
is originating sufficiently far from the active galactic nucleus of the Cloverleaf.

}
\maketitle

\section{Introduction}

There is evidence for solar or super-solar metallicities in the circumnuclear environments 
of quasars out to redshifts $z$$>$4 (e.g., Hamann \& Ferland 1999; Kurk et al. 2007; 
Jiang et al. 2007; Juarez et al. 2009; Matsuoka et al. 2009). This evidence, mainly 
from optical lines, is supported by millimeter detections of CO and dust in high-redshift 
sources, indicating rapid metal enrichment due to starbursts in the circumnuclear regions 
of at least some galaxies in the early universe (e.g., Solomon \& Vanden Bout 2005). This 
enrichment, however, might apply mainly to atomic nuclei that are synthesized 
in short-lived massive stars, and not so much to ``secondary'' nuclei like $^{13}$C 
that are thought to be mainly synthesized in longer-lived, less-massive stars (but
see, e.g., Hamann et al. 2002 for the mainly secondary element nitrogen).

In the local universe, $^{12}$C/$^{13}$C abundance ratios are sometimes considered to be a 
diagnostic of deep stellar mixing and a measure of ``primary'' vs.\ ``secondary'' nuclear 
processing (e.g., Wilson \& Rood 1994). While $^{12}$C is produced by He burning on rapid
time scales in massive stars, $^{13}$C is mainly synthesized by CNO processing of $^{12}$C 
seed nuclei from earlier stellar generations. This processing occurs more slowly, during 
the red giant phase in low- and intermediate-mass stars or novae. The $^{12}$C/$^{13}$C 
ratio may therefore depend on the nucleosynthesis history. It could be much higher in 
high-$z$ galaxies that are too young to have synthesized large amounts of secondary nuclei 
like $^{13}$C.

At optical, near-IR, and UV wavelengths it is difficult to discriminate between an element's 
isotopes because their atomic lines are blended (e.g., Levshakov et al. 2006). The prospects 
are better with radio lines from isotopic substitutions in molecules, which are well separated 
by a few percent of their rest frequency from the main species. This separation allows both 
the main and rare species to be easily identified, and to be observable with the same radio 
receivers and spectrometers.

The Cloverleaf Quasar (H1413+117), partly because of amplification by gravitational lensing, 
is a high-$z$ source with exceptional peak flux densities in $^{12}$C$^{16}$O (hereafter 
$^{12}$CO; see Appendix~2 of Solomon \& Vanden Bout 2005). This source is therefore 
one of the best candidates to search for $^{13}$C$^{16}$O (hereafter $^{13}$CO) to try to test 
models of ``chemical'' evolution over a Hubble time. In this paper we report on a search for 
$^{13}$CO(3--2) emission in the Cloverleaf at $z$=2.5579, when the universe was 2.5\,Gyr old.

\section{Observations}

The measurements were made with the IRAM Interferometer on Plateau de
Bure, France, in July, August, and September 2008, with 5 antennas in
the compact D-configuration (maximum baseline 97\,m) and the new 
dual-polarization receivers. The receiver and system single-sideband 
temperatures were 40 and 100\,K, respectively. The spectrometers covered 1\,GHz 
in each polarization, and the raw spectral resolution was 2.5\,MHz, 
or 8.1\,km\,s$^{-1}$. The data were binned to various spectral resolutions; 
in this paper we present data binned in 19$\times$160\,km\,s$^{-1}$ channels, 
covering a range of 3040\,km\,s$^{-1}$, with a noise of 0.22\,mJy\,beam$^{-1}$ 
(1$\sigma$) in each channel. The naturally-weighted synthesized beam
was 5\ffas6$\times$4\ffas8 at p.a.  62$^{\circ}$. Because the four CO
spots of the lensed Cloverleaf image are spread over 1\ffas7, we included 
more of the total flux by applying to the $u,v$ data a Gaussian taper 
that fell to $1/e$ at a radius of 100\,m. The slightly broadened beam 
then became 6\ffas1$\times$5\ffas4, and the noise in the individual 
channels is 0.23\,mJy\,beam$^{-1}$.

\begin{figure}[t]
\vspace{-0.0cm}
\centering
\includegraphics[angle=0,width=8.5cm]{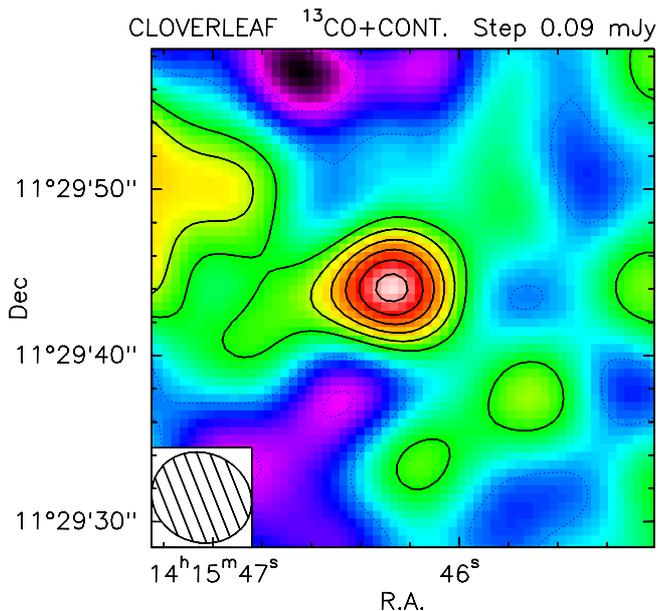}
\vspace{-0.0cm}
\caption{Contour map of continuum plus $^{13}$CO $J$=3$\rightarrow$2 emission, covering 
the central 960\,km\,s$^{-1}$ toward the Cloverleaf QSO. The beam is 6\ffas1$\times$5\ffas4 
(lower left) and the contour step is 0.09\,mJy (1\,$\sigma$). The peak value and the 
spatially-integrated intensity in the central source are both 0.6\,mJy.  
\label{fig1}}
\end{figure}
\begin{figure}[t]
\vspace{-0.0cm}
\centering
\includegraphics[angle=0,width=8.5cm]{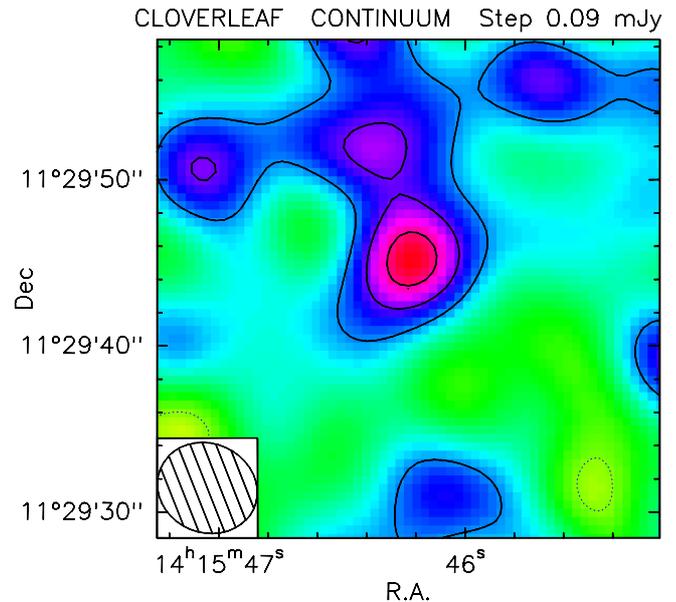}
\caption{Contour map of the 3.2\,mm continuum emission, covering 2080\,km\,s$^{-1}$
in the off-line channels. The beam is 6\ffas1$\times$5\ffas4 (lower left) and the 
contour step is 0.09\,mJy (1.5\,$\sigma$). The peak and the spatially-integrated 
flux density of the central source are both 0.3\,mJy. Combining this figure with the 
previous one, we conclude that in the 960\,\kms \ band centered on the $^{13}$CO 
$J$=3$\rightarrow$2 line, line and continuum each contribute half of the total 
flux density.  
\label{fig2}}
\end{figure}

\section{Results}

Figures~1 through 3 show the data, and Table~1 summarizes the results.
In the integrated line + continuum map (Fig.~1), the peak position
(Table~1) agrees well with the centroid of previous high-resolution
interferometer maps of the source (e.g., Alloin et al. 1997; Yun et
al. 1997; Kneib et al. 1998).  At 93\,GHz, the expected continuum is
0.30--0.35\,mJy (from Fig.~3 of Wei{\ss} et al. 2003 and the power-law 
given in Bradford et al. 2009) and a map in the 13 off-line channels 
at the positive and negative velocity ends of our spectra indeed yields 
a continuum flux of 0.3$\pm$0.1\,mJy (Fig.~2). This continuum adds to 
the line signal, and for this reason, the line appears much broader than 
the $\sim 430$ \,\kms\ widths of the $^{12}$CO and [C{\sc i}] lines 
(Wei{\ss} et al. 2003). The observed line has a low signal-to-noise 
ratio, which prevents a clear distinction between line and continuum, 
and does not allow us to constrain the line shape. Above the 0.3\,mJy 
continuum, a Gaussian fit yields an integrated line flux of 
(0.3$\pm$0.1)\,Jy\,km\,s$^{-1}$ (Fig.~3, see also the much higher upper 
limit given by Barvainis et al. 1997, their Table~1). An alternative 
Gaussian fit, with the line width fixed to the width of the $^{12}$CO 
line, yields a peak line flux density of (0.44$\pm$0.12)\,mJy\,beam$^{-1}$, 
and the same integrated line flux as the fit shown in Fig.~3. This 
integrated flux, corrected for frequency squared, leads to a 
$^{12}$CO/$^{13}$CO $J$=3$\rightarrow$2 line luminosity ratio 
($=$ brightness temperature ratio) of 40$^{+25}_{-8}$ (Table~1). 
This value is conservative. With the line width fixed to the width
of the $^{12}$CO line and the actual peak flux density of order 0.35\,mJy, 
the ratio would become $\sim$75.

\begin{table}
\label{tab1}
\begin{threeparttable}
\caption{$^{13}$CO(3--2) Observations and results.}
\begin{tabular}{lc}
\hline
{Parameter}&{$^{13}$CO(3--2)}\\
\hline
\multicolumn{2}{l}{{\it Observed CO(3--2) quantities:}}
\\
R.A. (J2000)              &14$^{\rm h}$ 15$^{\rm m}$ 46\ffs28 $\pm$ 0\ffs03\\
Dec. (J2000)              &+11$^{\circ}$ 29$'$ 44\ffas0 $\pm$ 0\ffas4 \\
Center frequency (GHz)    &92.91816         \\
Redshift (LSR) $^{\rm a)}$   &$2.55784\pm 0.00003$   \\
Continuum flux density (mJy) &$0.3\pm 0.1$ \\
Integrated $^{13}$CO flux (Jy\,\kms ) &$0.3\pm 0.1$    \\
\\
\multicolumn{2}{l}{{\it Derived CO(3--2) quantities:}}
\\
$L^\prime$($^{13}$CO (\Kkmspc )$^{\rm c)}$     &($1.1\pm 0.3$)$\times 10^{10}$  \\
$L^\prime$($^{12}$CO (\Kkmspc )$^{\rm c)}$     &($45.9\pm 3$)$\times 10^{10}$  \\
$L^\prime$ ratio $^{12}$CO/$^{13}$CO(3--2)   &40$^{+25}_{-8}$    \\
\hline
\end{tabular}
\begin{tablenotes}
\item[a)] adopted from $^{12}$CO (Weiss et al. 2003).
\item[b)] in a beam of 6\ffas1$\times$5\ffas4.
\item[c)] This is the lens-amplified value for a luminosity distance of 
$D_L$ = 21.28\,Gpc ($H_0 =$71\,\kms\,Mpc$^{-1}$, $\Omega_m =$ 0.27, 
$\Omega_{\rm vac} =$ 0.73) and an angular diameter distance of $D_A$ = 1.682\,Gpc; 
linear scale: 1$'' \leftrightarrow 8152$\,pc (Wright 2006).

\end{tablenotes}
\end{threeparttable}
\end{table}

\section{Large velocity gradient model calculations}

$^{12}$CO lines have higher optical depths than those of $^{13}$CO.
Therefore, the measured $^{12}$CO/$^{13}$CO line intensity ratio
(Sect.\,3) is a lower limit to the $^{12}$CO/$^{13}$CO abundance ratio.
To further constrain the $^{12}$CO/$^{13}$CO abundance ratio of the
Cloverleaf QSO, Table~2 provides flux densities and brightness
temperatures of seven $^{12}$CO transitions. To simulate these values,
a Large Velocity gradient (LVG) model was used with collision rates 
from Flower (2001), a cosmic microwave background of 9.7\,K, and an 
ortho-to-para H$_2$ abundance ratio of three (e.g., Wei{\ss} et al. 2005, 
2007; Riechers et al. 2006b).  The latter is, however, not critical 
for this study.

\begin{figure*}[t]
\vspace{-0.0cm}
\centering
\resizebox{17.8cm}{!}{\rotatebox[origin=br]{-90.00}{\includegraphics{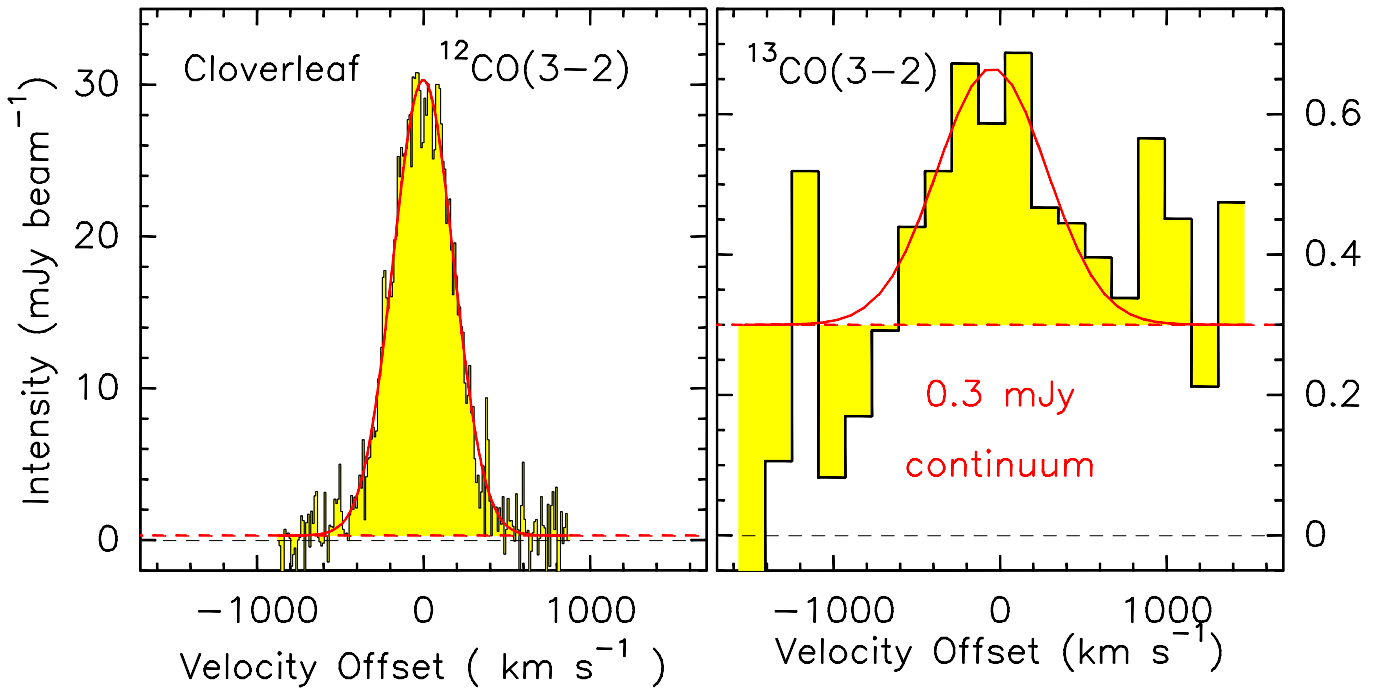}}}
\vspace{-0.0cm}
\caption{CO $J$=3$\rightarrow$2 from the Cloverleaf QSO, measured with the IRAM
interferometer. {\it Left:} \CO32 profile in 10\,\kms\ channels from 
Weiss et al. (2003). Velocity offsets are relative to 97.1928\,GHz
{\it Right:} \13CO32 profile in 160\,\kms\ channels, from this paper.
Velocity offsets are relative to 92.91816\,GHz ($z$ = 2.55784).
The red curves show Gaussian fits above a continuum of 0.3\,mJy. 
\label{fig3}}
\end{figure*}

We calculated a grid for $^{12}$CO/$^{13}$CO with kinetic temperatures
of 30--100\,K and $^{12}$CO fractional abundances per velocity interval
of [$^{12}$CO]/([H$_2$](d$v$/d$r$)) = 10$^{-4...-7}$\,pc\,(km\,s$^{-1}$)$^{-1}$. 
Accounting for possible effects of cloud structure, not only a spherical 
but also a plan-parallel cloud morphology was considered, with escape
probabilities $\beta_{\rm spherical}$ = (1--e$^{-\tau}$)/$\tau$ and
$\beta_{\rm plan-parallel}$ = (1--3e$^{-\tau}$)/(3$\tau$), respectively
($\tau$: optical depth). Resulting $^{12}$CO/$^{13}$CO abundance 
ratios reproducing the six measured $^{12}$CO line intensity ratios 
(Table~2) are given in Figs.~4 and 5 together with reduced $\chi^2$ 
($\chi^2_{\rm red}$) values of the best fit. We adopted a 1$\sigma$ error 
of 15\% for each fitted brightness temperature ratio. The dependence of the 
resulting $^{12}$CO/$^{13}$CO ratios on cloud morphology is caused by the different 
escape probabilities, related to $\tau$ in the case of a spherical and to 
3$\tau$ in the case of a plan-parallel cloud geometry. Therefore, a required 
amount of excitation through photon trapping is reached at lower $^{12}$CO 
optical depths in the case of a plan-parallel morphology, resulting in smaller 
$^{12}$CO/$^{13}$CO abundance ratios. 

\begin{table}
\label{tab2}
\begin{threeparttable}
\caption[]{CO line ratios in the Cloverleaf.}
\begin{flushleft}
\begin{tabular}{cccc}
\hline
Line        & Integrated         & $T_b$ ratio$^{\rm a}$     &  Reference$^{\rm b}$ \\
            & line flux          & to                        &        \\
            & (Jy\,km\,s$^{-1}$) & $^{12}$CO(3--2)           &        \\
\hline
            &                    &                           &        \\
CO(3--2)    & 13.2$\pm$2.0       & 1.00$\pm$0.15             &   1    \\
CO(4--3)    & 21.1$\pm$3.2       & 0.90$\pm$0.13             &   2    \\
CO(5--4)    & 24.0$\pm$3.6       & 0.65$\pm$0.09             &   2    \\
CO(6--5)    & 37.0$\pm$5.6       & 0.70$\pm$0.10             &   3    \\
CO(7--6)    & 45.3$\pm$6.8       & 0.63$\pm$0.09             &   3    \\
CO(8--7)    & 51.4$\pm$7.7       & 0.55$\pm$0.08             &   3    \\
CO(9--8)    & 41.8$\pm$6.3       & 0.35$\pm$0.05             &   3    \\
            &                    &            \\
$^{13}$CO(3--2) &0.3$\pm$0.1     & 0.025$^{+0.006}_{-0.009}$ &   4    \\
            &                    &                                 \\
\hline
\end{tabular}
\begin{tablenotes}
\item[a)] If all lines have the same area filling factor.
Adopted 1$\sigma$ errors are $\pm$10\% for the flux densities and $\pm$15\% for the 
brightness temperature ratios.
\item[b)] (1) Wei{\ss} et al. (2003); (2)  Barvainis et al. (1997);
(3) Bradford et al. (2009); (4) this paper.
\end{tablenotes}
\end{flushleft}
\end{threeparttable}
\end{table}

The $\chi^2_{\rm red}$ values displayed in Figs.~4 and 5 indicate that the 
CO data can be fitted by a single molecular gas component (cf. Bradford et al. 2009). 
All calculations are also consistent with the (not very stringent) upper limits 
for the $^{13}$CO $J$ = 7$\rightarrow$6 and 8$\rightarrow$7 flux densities from
Bradford et al. (2009). At first sight, the figures do not strongly reduce the 
permitted parameter space, providing $\chi^2_{\rm red}$ values of order 1.25--2. 
In the upper left corners of each figure, however, the $\chi^2_{\rm red}$ values 
rise significantly, becoming too large to provide credible solutions. As a consequence, 
the overall $^{12}$CO/$^{13}$CO abundance ratio appears to be $>$100 in the Cloverleaf 
QSO. There exist further constraints: (1) $T_{\rm kin}$ $<$ 30\,K is prohibitive 
because of the temperatures determined from C{\sc i} ($\sim$30\,K) and the dust 
($\sim$50\,K, Wei{\ss} et al. 2003). Furthermore, such low temperatures would require 
extreme CO column densities to raise photon trapping to such levels that the emission 
from the higher $J$ transitions could be reproduced. (2) $T_{\rm kin}$ $>$ 50\,K is 
also not likely because of the temperature deduced from [C{\sc i}] and the close 
association of CO and C{\sc i}, which appears to be independent of the environment 
(e.g., Ikeda et al. 2002; Zhang et al. 2007). 

\begin{figure}[t]
\vspace{-4.0cm}
\centering
\resizebox{17.8cm}{!}{\rotatebox[origin=br]{-90.00}{\includegraphics{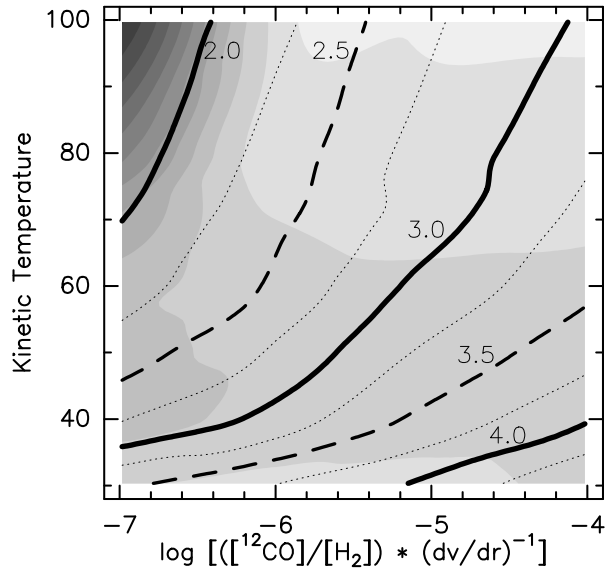}}}
\vspace{-0.0cm}
\caption{Results from large velocity gradient (LVG) radiative transfer
calculations using a spherical cloud model to simulate the line intensity ratios
given in Table~2. The common logarithm of the $^{12}$CO/$^{13}$CO abundance ratio 
is shown as a function of kinetic temperature in units of Kelvin and 
of fractional abundance in units of pc\,(km\,s$^{-1}$)$^{-1}$ for a 
$^{12}$CO/$^{13}$CO $J$=3$\rightarrow$2 line intensity ratio of 40. Resulting 
reduced $\chi^2$ values ($\chi^2_{\rm red}$) for the simulation of the four 
$^{12}$CO lines given in Table~2 are shaded. Lightest grey: 1.00 
$\leq$ $\chi^2_{\rm red}$ $<$ 1.25, darker shades of grey at 1.25, 1.50, 
1.75 to 4.00 with an increment of 0.25. The maximum value in the upper left 
corner is $\chi^2_{\rm red}$ = 4.07. 
\label{fig4}}
\end{figure}

\begin{figure}[t]
\vspace{-4.0cm}
\centering
\resizebox{17.8cm}{!}{\rotatebox[origin=br]{-90.00}{\includegraphics{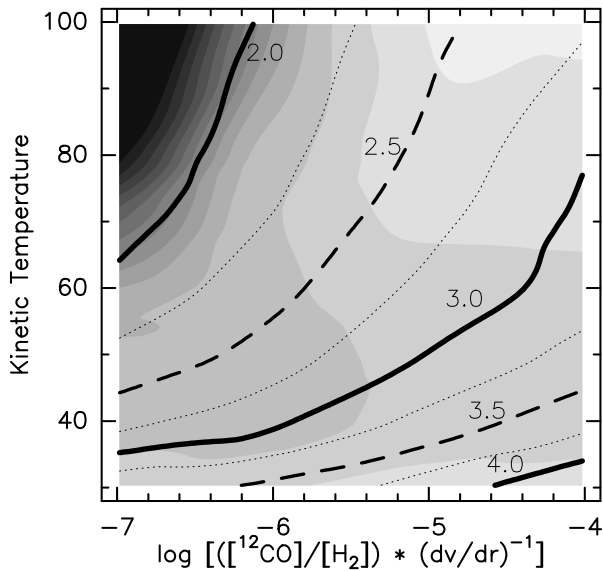}}}
\vspace{-0.0cm}
\caption{Same as Fig.~4, but for a plan-parallel cloud geometry.
Resulting $\chi^2_{\rm red}$ values for the simulation of the line intensity 
ratios given in Table~2 are shaded. Lightest grey: 1.00 $\leq$ $\chi^2$ $<$ 1.25, 
darker shades of grey at 1.25, 1.50, 1.75 to 4.00 with an increment of 0.25. 
The maximum value in the upper left corner is $\chi^2_{\rm red}$ = 6.31.
\label{fig5}}
\end{figure}

For 30$\leq$T$_{\rm kin}$$\leq$50\,K and [$^{12}$CO]/([H$_2$](d$v$/d$r$)) = 
10$^{-7}$\,pc\,(km\,s$^{-1}$)$^{-1}$, we obtain $^{12}$CO/$^{13}$CO abundance 
ratios in the range 200--3000 (Figs.~4 and 5). However, such a low 
fractional abundance per velocity interval can be firmly excluded. With 
[C{\sc i}]/[H$_2$] reaching values in agreement with those of the local 
galactic disk (Wei{\ss} et al. 2005), the [$^{12}$CO]/[H$_2$] abundance ratio 
should be of order 10$^{-4}$ (e.g., Frerking et al. 1982). The resulting 
velocity gradient of d$v$/d$r$ = 10$^3$\,km\,s$^{-1}$\,pc$^{-1}$ would be 
far too large in view of the measured line width (e.g., Wei{\ss} et al. 2003) 
and the kinetic temperature such extreme conditions would induce (e.g.,
Wiklind \& Henkel 2001). A velocity gradient of a few km\,$^{-1}$\,pc$^{-1}$ 
is more realistic as, e.g., obtained from clouds in virial equilibrium
for densities of order 10$^4$\,cm$^{-3}$ (Goldsmith 2001; his equation 2).
Such densities are commonly derived for high $z$ sources (e.g., Wei{\ss} et al. 
2005, 2007). For diffuse clouds, velocity gradients should be larger
(e.g., Papadopoulos et al. 2010). Bradford et al. (2009) suggest that in the 
Cloverleaf the velocity dispersion may exceed the virial requirement by at 
least an order of magnitude. Therefore the best choice may be [$^{12}$CO]/([H$_2$](d$v$/d$r$)) 
= 10$^{-5...-6}$\,pc\,(km\,s$^{-1}$)$^{-1}$ (for the higher value see, e.g., 
Riechers et al. 2006b; Wei{\ss} et al. 2007) to simultaneously fit the observed CO 
transitions from $J$=1$\rightarrow$0 up to 11$\rightarrow$10. Depending on the 
adopted kinetic temperature (30--50\,K) and cloud morphology, and irrespective 
of the optimal [$^{12}$CO]/([H$_2$](d$v$/d$r$)) value (as long as it is in 
the wide range displayed by Figs.\,\ref{fig4} and \ref{fig5}) we then find 
a $^{12}$CO/$^{13}$CO abundance ratio in the range 300--10000. 
In the following we will discuss whether this estimate can be realistic.

\section{Discussion}

In order to further evaluate our observational result, we have to discuss 
the correlation between molecular $^{12}$CO/$^{13}$CO and atomic 
$^{12}$C/$^{13}$C abundance ratios and to summarize relevant observational 
data from low-redshift galaxies which are, like the Cloverleaf, ultraluminous 
in the infrared. Finally, we will address some fundamental problems, which are 
related to the still poorly known morphology of the gas surrounding the 
Cloverleaf QSO.

\subsection{Chemical fractionation and isotope selective photodissociation}

Observed isotope ratios may be affected by fractionation. The
$^{12}$CO/$^{13}$CO abundance ratio is likely influenced by the
reaction
$$ 
^{13}{\rm C}^+ + ^{12}{\rm CO} \rightarrow\ ^{12}{\rm C}^+ +
^{13}{\rm CO} + \Delta E_{\rm 35K}
$$ 
(Watson et al. 1976). The process enhances $^{13}$CO relative to
$^{12}$CO in the more diffuse C$^+$ rich parts of molecular
clouds. This may be compensated by isotope selective
photodissociation. $^{12}$CO and $^{13}$CO need similar amounts of
self-shielding to survive in a hostile interstellar environment. This
favors the more abundant isotopologue (e.g., Sheffer et al. 2007).

For the Galaxy, such effects can be quantified. Milam et al. (2005)
summarized $^{12}$C/$^{13}$C ratios from the galactic disk, obtained
with the three molecules CO, CN, and H$_2$CO. These molecular species
are synthesized by quite different chemical reactions. The good
agreement between their $^{12}$C/$^{13}$C ratios and a lack of
correlation with kinetic temperature suggests that chemical
fractionation as well as isotope selective photodissociation do not
greatly affect the determined isotope ratios.

Whether this result is also valid in the case of the Cloverleaf QSO
may not be obvious at first sight. The ultraviolet radiation field in
the vicinity of the quasar might be exceptionally strong, favoring
$^{12}$CO over $^{13}$CO and thus leading to an enhanced molecular
abundance ratio with respect to $^{12}$C/$^{13}$C. However, such a
scenario is not likely.  Firstly, most of the galactic data were
obtained toward prominent sites of massive star formation, where the
UV radiation field is also exceptionally intense. Secondly, judging 
from C{\sc i}, in the Cloverleaf the excitation of the molecular gas 
is intermediate between conditions found for the starburst galaxy 
M\,82 ($T_{\rm ex,CI}$ $\sim$50\,K) and the central region of the 
Milky Way ($T_{\rm ex,CI}$ $\sim$ 22\,K) (Stutzki et al. 
1997; Wei{\ss} et al. 2003). Thirdly, polycyclic aromatic hydrocarbon 
(PAH) features are as strong as expected with respect to the far 
infrared luminosity when compared with more nearby ultraluminous
star-forming galaxies, favoring ``normal'' conditions and a 
predominantly starburst nature of the Cloverleaf's huge FIR emission 
(Lutz et al.  2007). Finally, the CO emission from the Cloverleaf 
appears to be more extended than the effective radius out to which 
the quasar could dominate the UV field. 

Modeling both the source and the lens of the Cloverleaf QSO, 
Venturini \& Solomon (2003) find a characteristic radius of $r$ 
$\sim$ 800\,pc for the CO $J$=7--6 line, which is higher excited and 
thus possibly less widespread than the $J$=3$\rightarrow$2 transition
considered here. If the Cloverleaf's intrinsic far infrared luminosity 
($L_{\rm FIR}$ $\sim$ 5$\times$10$^{12}$\,L$_{\odot}$, Lutz et al.  
2007) would entirely originate from 6.2--13.6\,eV photons emitted by 
the active nucleus, we would obtain, at a radius of 800\,pc, a UV 
photon illumination of $\chi$ $\sim$ 10$^5$\,$\chi_0$ with respect 
to the local galactic radiation field, $\chi_0$ =
2$\times$10$^{-4}$\,\,erg\,cm$^{-2}$\,s$^{-1}$\,sr$^{-1}$ (see Draine
1978). The Cloverleaf QSO is a Broad Absorption Line (BAL) quasar
which permits at least a partial view onto its nuclear
engine. Therefore, taking the Cloverleaf's UV luminosity from Fig.\,1
of Barvainis et al. (1995) and accounting for a gravitational
amplification by a factor of 11 (Solomon \& Vanden Bout 2005), we
obtain accordingly $\chi$ $\sim$ 2.5$\times$10$^4$\,$\chi_0$. Both
$\chi$ values are consistent with those encountered in prominent
galactic sites of massive star formation and may be upper limits 
if the Cloverleaf posseses a self-shielding rotating disk. To summarize,
physical conditions in the Cloverleaf host galaxy appear to be sufficiently
normal so that the $^{12}$C/$^{13}$C isotope ratio should not strongly 
deviate from the $^{12}$CO/$^{13}$CO molecular abundance ratio.

\subsection{$^{12}$CO/$^{13}$CO ratios in $z$$<$1 galaxies}

{\it In our Galaxy}, the $^{12}$CO/$^{13}$CO line intensity ratios from
molecular clouds are typically about 5, probably corresponding to true
$^{12}$C/$^{13}$C abundance ratios of $\sim$25 in the galactic Center, 
$\sim$50 in the inner galactic disk and the LMC, $\sim$70 at the Sun's 
galactocentric radius, and  $\ga$100 in the outer Galaxy. The solar system 
ratio of 89 may have been typical of the galactic disk at the Sun's 
galactocentric radius 4.6\,Gyr ago (e.g., Wilson \& Rood 1994; Wouterloot 
\& Brand 1996; Wang et al. 2009). Within the framework of ``biased infall'',
where the galactic disk developed from inside out (Chiappini et al. 2001), 
there {\it might} be a future chance to use $^{12}$C/$^{13}$C ratios as a 
chronometer for nucleosynthesis. 

{\it In nearby galaxies}, the $^{12}$CO/$^{13}$CO line intensity
ratios are usually measured in the $J$=1--0 line and have typical
values of $\sim$10. They are higher than the values for individual
molecular clouds in the Galaxy because they are mostly observed with 
larger beams. These include not only the dense clouds, where both 
species are (almost) optically thick, but also the molecular intercloud 
medium, where $^{13}$CO is optically thin. Like the better-resolved 
CO line ratios in our Galaxy, the ratios in nearby galaxies probably 
correspond to true $^{12}$C/$^{13}$C abundance ratios between 40 and 
90 (e.g., Henkel et al. 1993).

In a presumably ``normal'' spiral {\it galaxy at redshift 0.89}, in 
the lens of the background source PKS\,1830-211, Wiklind \& Combes 
(1998),  Menten et al. (1999), and  Muller et al. (2006) derive, 
from the optically thin wings of the absorption lines of HCO$^+$, 
HCN, and HNC, a $^{12}$C/$^{13}$C abundance ratio of 27$\pm$2. 
Apparently, even at an age of the universe of $\sim$6.5\,Gyr, it 
appears that $^{13}$C is as abundant with respect to $^{12}$C as 
in the center of our Galaxy at the present epoch.

{\it Some low-redshift (ultra)luminous infrared galaxies} ((U)LIRGs), 
however, show peculiarities, which may be relevant to the Cloverleaf.
Local (U)LIRGs are known to reveal $^{12}$CO/$^{13}$CO $J$= 1$\rightarrow$0
line intensity ratios which tend to be higher than the canonical value 
of 10 for ``normal'' galaxies (see, e.g., Aalto et al. 1991; Casoli et 
al. 1992; Henkel \& Mauersberger 1993). According to Taniguchi \& Ohyama 
(1998), there is a tight correlation between $L$($^{12}$CO 
$J$=1$\rightarrow$0) and $L_{\rm FIR}$. However, when comparing
``normal'' galaxies with those with a high $^{12}$CO/$^{13}$CO 
$J$=1$\rightarrow$0 ratio, the $^{13}$CO
luminosities show a deficiency by an average factor of $\sim$3, 
This $^{13}$CO deficiency is readily explained by metallicity 
gradients in the progenitor galaxies and strong interaction- or 
merger-induced inflow of gas into the luminous cores (e.g., 
Rupke et al. 2008). Apparently, for ultraluminous galaxies the common
luminosity - metallicity correlation is not valid. Ultraluminous 
galaxies are characterized by a lower metallicity, likely
yielding higher $^{12}$C/$^{13}$C isotope ratios. In the early 
universe, gas from outside the cores of the merging progenitors 
may have been particularly metal poor, leading to extreme carbon 
isotope ratios. 

For $T_{\rm kin}$ $\ga$ 20\,K, the $^{12}$CO $J$=3$\rightarrow$2 line 
is more opaque, typically by a factor of 3, than the corresponding 
1$\rightarrow$0 line. Thus our conservatively estimated $J$=3--2 
$^{12}$CO/$^{13}$CO line intensity ratio of $\ga$40$^{+25}_{-8}$ 
corresponds to a 1$\rightarrow$0 ratio well in excess of 40. So far, 
only few $^{12}$CO/$^{13}$CO $J$=3$\rightarrow$2 line ratios have been
measured in luminous mergers of low redshift. Greve et al. (2009) find 
8$\pm$2 for the ULIRG Arp~220 and $\ga$30 for the LIRG NGC~6240.
The latter value {\it might} be consistent with that of the Cloverleaf.

\subsection{Are there alternatives to a $^{13}$C deficiency 
in the Cloverleaf?}

Sects.\,5.1 and 5.2 suggest, that our measured $^{12}$CO/$^{13}$CO
line intensity ratio (or its lower limit) require a significant $^{13}$C
deficiency in the Cloverleaf. Are there caveats we may have overlooked 
when reaching this conclusion?

If the bulk of the CO emission would not arise, as suggested
by Venturini \& Solomon (2003), from a molecular disk but from a large
scale outflow, such gas would not be in virial equilibrium and could
arise predominantly from a diffuse gas phase. While this would yield 
(within the LVG approach) a higher velocity gradient and a lower 
[$^{12}$CO]/([H$_2$](d$v$/d$r$)) value than what is needed for 
virialized clouds, required densities would then be well in excess
of 10$^4$\,cm$^{-3}$, in contradiction with our assumption of 
predominantly diffuse gas. Furthermore, as long as $T_{\rm kin}$ remains 
moderate ($\la$50\,K; see Figs.\,\ref{fig4} and \ref{fig5}), $^{12}$C/$^{13}$C 
ratios remain larger than those encountered in the galactic disk (Sect.\,5.1).  

Following White (1977), radiative transfer models with simple geometry,
either based on microturbulence or on systematic motions, lead to peak
and integrated intensities which agree within the differences (up to
a factor of three) caused by an uncertain cloud geometry. A full 
3-D model of a rotating circumnuclear disk, computing the radiative 
transfer through many lines of sight, calculating the LVG level 
populations within each pixel of the simulated source, and also 
including continuum radiation from dust (e.g., Downes \& Solomon 
1998) may be worth doing. In the Cloverleaf, however, the 
distribution of the molecular gas is still poorly known. 

A large $^{12}$C/$^{13}$C ratio, implying an underabundance of 
$^{13}$C, appears to be in direct conflict with optical data. As 
already mentioned in Sect.\,1, solar or super-solar metallicities 
are common in quasars up to high redshifts. This does not only refer 
to so-called ``$\alpha$-elements'' being rapidly synthesized in 
short-lived massive stars but also to iron (e.g., Iwamuro et al. 
2004; Kurk et al. 2007; Sameshima et al. 2009), carbon (e.g., Jiang 
et al. 2007; Juarez et al. 2009), and, even more importantly, nitrogen 
(Hamann \& Ferland 1999; De Breuck et al. 2000; Vernet et al. 2001; 
Hamann et al. 2002; Nagao et al. 2006; Matsuoka et al. 2009), with
$^{14}$N being mainly a secondary nucleus produced by CNO burning 
just like $^{13}$C. A possible explanation for the contradictory results
obtained at or near optical wavelengths and the microwave data presented 
here may be different locations. It is well possible that mainly 
secondary nuclei like $^{13}$C and $^{14}$N are enriched close to the 
quasar, in the Broad and Narrow Line Regions and in outflows originating 
from the active galactic nucleus (AGN). However, CO $J$=7$\rightarrow$6 
may arise hundreds of pc away from the AGN (Venturini \& Solomon 2003) 
and some of the $J$=3$\rightarrow$2 photons may be emitted from locations 
even farther away.

There exists, however, also the possibility that our measured high 
$^{12}$CO/$^{13}$CO luminosity ratio is misleading and does {\it not}
imply a large $^{12}$C/$^{13}$C ratio. As a consequence of different 
optical depths, $^{12}$CO lines are almost thermalized and 
are characterized by excitation temperatures well above the level of the 
cosmic microwave background even at $z$=2.5. $^{13}$CO is less thermalized. 
In our best fitting models, its $J$=3$\rightarrow$2 excitation temperature 
lies in the range 20--30\,K. This is significantly above the 9.7\,K of the 
CMB. However, an extreme (and therefore unlikely) enhancement of the 
background level by dust radiation could reduce the contrast between 
line and background for $^{13}$CO far more efficiently than for $^{12}$CO 
(see Papadopoulos et al. 2010 for the case of Arp\,220), thus
establishing an apparent $^{13}$CO deficiency.

\section{Outlook}

Molecular lines from galaxies in the distant universe have the
potential to reveal the contribution of early stellar generations
to the enrichment of the interstellar medium. Our data from the
$z$ = 2.5 Cloverleaf QSO are a first step toward studying the 
isotopic composition of such gas in the distant past. Our data 
indicate, not unexpectedly, a strong deficiency of $^{13}$C with respect 
to $^{12}$C in the host galaxy. However, the weakness of the tentatively 
detected line, the limited number of observed transitions, the poorly
constrained source morphology, and the potential influence of an enhanced 
submillimeter radiation background do not yet allow us to derive a definite 
$^{12}$C/$^{13}$C isotope ratio. Significant progress in this field either 
requires the detection of stronger sources or the higher instrumental 
sensitivity of the Atacama Large Millimeter Array (ALMA), which will allow 
us to study the isotopes of C, N, and O in a number of highly redshifted 
targets. Toward the Cloverleaf, the main isotopologes of HCN, HCO$^+$, and 
CN (Solomon et al. 2003; Riechers et al. 2006a, 2007) have already been 
detected.

\begin{acknowledgements}
We wish to thank P.P. Papadopoulos, D. Riquelme, S. Veilleux, and an
anonymous referee for helpful discussions on ULIRGs and chemical evolution
and/or a critical reading of the manuscript. This paper is based on observations 
taken with the IRAM Plateau de Bure Interferometer. IRAM is supported by 
INSU/CNRS (France), the MPG (Germany), and the IGN (Spain). DR acknowledges 
support from NASA through Hubble Fellowship grant HST-HF-01212.01-A, awarded 
by the Space Telescope Science Institute, which is operated by AURA for NASA 
under contract NAS5-26555. This research has made use of NASA's Astrophysical 
Data System (ADS).
\end{acknowledgements}

\end{document}